
\input harvmac

\Title{\vbox{\hbox{HUTP-95/A014} }      }
{\vbox{\centerline{A Stringy Test of the Fate of the Conifold}}}
\bigskip
\bigskip

\centerline{Cumrun Vafa}
\medskip\centerline{Lyman Laboratory of Physics}
\centerline{Harvard University}
\centerline{Cambridge, MA 02138, USA}

\vskip 1in
By studying string loop corrections to superpotential
of type II strings compactified on Calabi-Yau threefolds
 we find a quantum stringy test and a confirmation
of a recent proposal of Strominger on the fate of the conifold
singularity.
We also propose a connection between the spectrum of Bogomolnyi
saturated solitons and one-loop string partition function
of $N=2$ topological strings.
\Date{5/95}

Through some recent exciting work in string theory
it has become clear that nonperturbative aspects of string
theory are within reach, at least for certain string compactifications.
In particular a clear picture is emerging for qualitative
nonperturbative aspects of string theory for
type II string compactifications on tori \ref\hu{
C.M. Hull and P.K.  Townsend, ``Unity of Superstring
Dualities'', QMW-94-30, R/94/33.}
\ref\wi{E. Witten, ``String Theory Dynamics in Various Dimensions'',
hep-th/9503124.
}, on $K3$ manifold \hu \wi \ref\more{A. Sen, ``String-String Duality
Conjecture in Six Dimensions and
Charged Solitonic Strings'', hep-th/9504027\semi
J. Harvey and A. Strominger, ``The
Heterotic String is a Soliton'', hep-th/9504047}\
and on Calabi-Yau threefolds \ref\strom{A. Strominger
, ``Massless Black Holes and Conifolds in String Theory'', hep-th/950490
.}\ref\stromet{
 B. Greene, D. Morrison and A. Strominger, ``Black Hole Condensation
and the Unification of String Vacua'', hep-th/9504145.}.
In this paper we present some evidence
based on string loop corrections in favor of the beautiful picture
advocated in \strom\ for non-perturbative aspects of
type II strings on Calabi-Yau manifolds.  An argument
in favor of the scenario suggested in \strom\ is the exciting
discovery \stromet\ that it allows smooth transitions
among Calabi-Yau manifolds, thus realizing a suggestion made
in \ref\confo{ P. Candelas, A.M. Dale, C.A. L\"utken, and R. Schimmrigk,
``Complete Intersection Calabi--Yau Manifolds'', Nucl. Phys. B298 (1988)
493--525.}\ref\conan{P.S. Green and T.  H\"ubsch,
``Possible Phase Transitions Among Calabi--Yau Compactifications'',
Phys. Rev. Lett. 61 (1988) 1163--1166;
``Connecting Moduli Spaces of Calabi--Yau Threefolds'', Comm. Math.
Phys. 119 (1988) 431--441.}\ref\conmo{P. Candelas, P.S. Green, and T.
H\"ubsch,
``Finite Distance Between Distinct Calabi--Yau Manifolds'',
Phys. Rev. Lett. 62 (1989) 1956--1959;
``Rolling Among Calabi--Yau Vacua'', Nucl. Phys.
 B330 (1990) 49--102.}\ref\ccon{ P. Candelas and X.C. de la Ossa,
``Comments on Conifolds'', Nucl. Phys. B342 (1990) 246--268.}.
It is amusing
that string loop corrections can be used to test some of these
non-perturbative aspects of string theory.
This is in accord
with a type IIA string one loop computation \ref\vw{C. Vafa and E. Witten,
in preparation.}\
which further supports the conjecture \hu\ref\vun{C. Vafa, unpublished.}\
that
type IIA compactified on $K3$ is dual to
toroidal compactification of heterotic string
down to 6 dimensions.

Let us recall the basic picture advocated in \strom\ for the
explanation of the conifold singularity in Calabi-Yau compactifications.
Type IIB strings compactified on a Calabi-Yau threefold leads to a
theory in 4d with $N=2$ supersymmetry\foot{It suffices to consider
type IIB, because type IIA is equivalent to the same theory on the
mirror Calabi-Yau.}. The massless string states
apart from the gravity multiplet include some $U(1)$ vector
multiplets and some hypermultiplets.  The number of vector
multiplets is equal to $h^{2,1}$ of the Calabi-Yau manifold $M$
and the number of hypermultiplets  is equal to $h^{1,1}+1$.
Since we have $N=2$ supersymmetry in 4 dimensions we can have
a central charge in the supersymmetry algebra.  This
could lead to BPS states which saturate the Bogomolnyi
bound and are thus stable.
  Such objects would have electric
or magnetic charge under the $U(1)$ gauge fields of the vector
multiplet.  However
there are no such states among the fundamental string excitations
as they are all neutral under the Ramond-Ramond $U(1)$ gauge fields.
It was suggested in \strom\ that by wrapping
certain threebrane solutions \ref\sol{G. Horowitz and A. Strominger, ``Black
Strings and
$p$-branes,'' Nucl. Phys. B360 (1991) 197.}\
 on three-cycles of
Calabi-Yau
one could construct extreme black hole solutions in four dimensions
which would carry $U(1)$ charges.  Assuming charge quantization
the allowed electric
and magnetic charges of such states would be in one to one correspondence
with $H_3(M,{\bf Z})$.  Let $C_i$ be an integral three-cycle of
$M$.  Let $\omega$ denote the holomorphic
three form on $M$. Then the central charge of the supersymmetry
algebra $W_i$ (up to normalization) for this black hole is given by
$$W_i=\int_{C_i}\omega$$
and the BPS mass is given by
\eqn\masf{m_i^2={\int_{C_i} \omega \cdot \int_{C_i} \bar \omega
\over \int_M \omega \wedge \bar \omega}}
(for a recent discussion on this see
\ref\cere{A. Ceresole, R. D'Auria, S. Ferrara and A. Van Proeyen,``
Duality Transformations in Supersymmetric Yang-Mills Theories
coupled to Supergravity'',hep-th/9502072.}).
The question of whether
all such classical solutions, and with what multiplicities,
lead to quantum states
is not clear.  In fact for the resolution of the conifold
singularity it was argued in \strom\ that not all of such
black hole states exist as fundamental fields in the quantum theory.
Consider a conifold point $p$ of the Calabi-Yau moduli space.
This is a point where by definition there exists a 3-cycle $C_p$
for which $W_p=\int_{C_p}\omega \rightarrow 0$ as we approach the conifold
point
(assuming a normalization of $\omega$ so that $\int
\omega \wedge\bar\omega
>0$).
A conifold point is a finite distance away from the interior of the moduli
space, and the moduli space
 has a curvature singularity at that point.  This would cause problem
for the predictive power of string theory.
This problem is more acute because one can show that the moduli
space metric is perturbatively uncorrected.  As has been elaborated
in \ref\beret{M. Bershadsky, S. Cecotti, H. Ooguri and C. Vafa,
``Kodaira-Spencer Theory of Gravity and Exact Results for Quantum
String Amplitudes'', Comm. Math. Phys. 165 (1994) 311.}\
the special geometry metric on the moduli
space of Calabi-Yau manifolds can be understood as an effective field
theory for the massless modes at the string tree level,
where one integrates out massive Kaluza-Klein states
of the Calabi-Yau.  This in particular means that
the origin of this tree level
singularity in string theory
can be understood by the interaction of the moduli
fields with massive string states that are becoming massless
at the conifold point.
This however is not the end of the story in the full
quantum theory of strings.  Even though we can understand
the classical origin of the singularity from the viewpoint
of nearly massless string states, such states are not stable
even at the tree level and thus
could hardly make sense in the full quantum theory.

We thus have two possibilities:  Either we can assume
these singularities are not there when we consider
non-perturbative string effects, or their origin
in the full theory can be understood by some other nearly
massless state. The latter was the viewpoint advocated in \strom.
This is somewhat similar to what happens for $N=1$
supersymmetric Yang-Mills in 4d with $G=SU(N_c)$ and with
$N_f=N_c+1$ where the moduli
space is classically singular and is uncorrected by quantum effects,
but the explanation of the singularity in the full theory is
different \ref\nsei{N. Seiberg, ``Exact Results on the Space
of Vacua of Four Dimensional SUSY Gauge Theories'', hep-th 9402044.},
where it is explained in terms of massless Baryons
and mesons.

 It was suggested in \strom\ that the metric on the moduli
space is not corrected even non-perturbatively and furthermore
near a conifold point $p$
the stable nearly massless field exists which
is a black hole state corresponding to the vanishing cycle $C_p$.
Such black hole states can be constructed by wrapping
the threebrane solution of \sol\ around
a three-cycle of the Calabi-Yau manifold.
For this to work it was crucial to assume that there are no
stable black holes corresponding to the cycles $nC_p$ for $|n|>1$.
The proposed black hole is an $N=2$ hypermultiplet.  Note
that if we start from the interior of the moduli space and approach
the conifold point $p$ the cycle that becomes a vanishing cycle will depend on
the path that we take to $p$.  It is natural then to consider
the totality of such vanishing cycles at each point on the moduli
space as they would be a natural candidate for the space of
stable black holes.  The totality of such vanishing cycles
can be obtained by considering the action of the monodromy group
of the moduli on a basis for the vanishing
three-cycles.  The monodromy group
is obtained by deleting codimension one singularities
of the moduli space (including orbifold points) and considering
the transformation of three-cycles around such singularities.
This will be a subgroup
 $H\subset Sp(h^{2,1}+1,{\bf Z})$.  Thus if $C_p$ is any one
of the vanishing cycles (and a primitive
three-cycle in the homology of $M$), we consider the space
$${\cal V}=\{ H\cdot C_p \}. $$
This is a candidate for at least a subspace of
stable black holes.
The cycles in ${\cal V}$ would all have to correspond
to stable black hole states were it not for the
fact that BPS states may appear or disappear when the
Bogomolnyi charges of two cycles are in the same
direction in the complex plane.  In particular suppose
$C_i$ and $C_j$ correspond to occupied black hole states,
with Bogomolnyi charges given by $W_i$ and $W_j$.  Then
if
$$W_i=\alpha W_j \qquad \alpha>0$$
where $\alpha$ is a positive real number, then a black hole
state corresponding to the cycle $C_i+C_j$ can appear or disappear.
This
 does happen in massive $N=2$ models in 2 dimensions which was discovered
and investigated in detail in \ref\cecv{S. Cecotti
and C. Vafa,``On Classification of N=2 Supersymmetric Theories'',
Comm. Math. Phys. 158 (1993) 569.}\ref\abcec{S. Cecotti,
P. Fendley, K. Intriligator and C. Vafa,`` A New Supersymmetric
Index'', Nucl. Phys. B386 (1992) 405.}.  In that case
the BPS states were related to the intersection
numbers of the vanishing cycles of the
Landau-Ginzburg singularity and the monodromy of the vanishing
cycle leads to a jump in the intersection numbers and thus
the number of BPS solitons.  It is tempting to speculate here
as well that the intersection numbers of the vanishing cycles
is related to the soliton number jumping.
Note that since the cycles $C_i$ are in dimension 3, their intersection
makes sense, just as was the case considered in \cecv .
Similar phenomenon of jumping of BPS solitons
were also postulated in \ref\sewi{N. Seiberg and E. Witten, ``Electromagnetic
Duality,
Monopole Condensation and Confinement in $N{=}2$ Supersymmetric
Yang-Mills Theory'',
Nucl. Phys.  B426 (1994) 19--52.}\
 in order to obtain a consistent
exact picture of 4 dimensional
$N=2$ theories.  Given that here we are also dealing with $N=2$
theories in 4 dimensions it is likely that we will also have
the jumping phenomenon.  Note that the jumping phenomenon is
inherently associated with having a complex Bogomolnyi
charge as is the case with $N=2$ supersymmetry. In such
cases if one wishes to move in the moduli of the theory
in certain cases one cannot avoid passing
through configurations with collinear Bogomolnyi charges
along a path joining two given points on the moduli.  This
is the reason for the possibility of having
jumping.  If we have $N>2$ we do not expect
such jumping phenomenon because the Bogomolnyi charge
lives in a bigger space, and in going from an initial point in the moduli
to a final point, we can avoid passing through collinear Bogomolnyi
charges.  Since whether we have jumping or not should depend
only on the end points on the moduli space this means we do not
have any jumping if $N>2$.

The tree level singularity near the conifold can be explained
by the existence of a nearly massless black hole state in the full theory.
Can we test this idea further using perturbative string amplitudes?
If we encounter quantum
string amplitudes which diverge as we approach the conifold
point, then there should be an alternative explanation of the singularity
from the viewpoint of the existence of massless black holes.
The main difficulty, apart from the practical question of computation
is that any computation we do may have correction at all loops and
even non-perturbatively.  We must thus consider objects which
are not corrected to all orders in perturbation theory and also
non-perturbatively.  In fact it has been argued that superpotential
terms considered upon compactification of type II strings on Calabi-Yau
manifolds are of this type and moreover
can be computed in string theory \beret \ref\nar{I. Antoniadis,
E. Gava, K.S. Narain and T.R. Taylor,``Topological Amplitudes in
String Theory'', Nucl. Phys. B413 (1994) 162.}\
by relating them to $N=2$ topological string amplitudes
on the Calabi-Yau manifold. In particular the genus $g$ topological
amplitude $F_g$ computes, as a function of the moduli, a correction
to the lagrangian.  These corrections involve
superpotential terms for the vector multiplet and
can be written conveniently in the superspace by
(with a similar formula for the hypermultiplets \nar )
\eqn\efac{\delta S=\int d^4\theta F_g W^{2g}\sim \int F_g R^2 T^{2g-2}+...}
where $W$ is the $N=2$ superfield associated to graviphoton field
strength,
$R$ is the Riemann tensor, and $T$ is the graviphoton field strength.
The contractions in the above terms are fixed by supersymmetry.
  The status of the non-perturbative
correction to these terms is on the same footing as the special geometry.
So it is reasonable to postulate that they do not get corrected even
non-perturbatively and so $F_g$ computes the full answer for the
correction of $W^{2g}$ to the theory.

The topological amplitudes $F_g$
 has been computed for some Calabi-Yau manifolds
for small $g$
 \beret \ref\holan{M. Bershadsky, S. Cecotti, H. Ooguri
and C. Vafa,``Holomorphic Anomalies in Topological
Field Theories'', Nucl. Phys. B405 (1993) 279.}.
Moreover the structure
of the singularity near the conifold point was determined
for all $g$ in \beret .
$F_1$ has in addition been studied for many more
examples including some with 2 parameter moduli space
\ref\multi{S. Hosono, A. Klemm, S. Theisen and S.-T. Yau,
``Mirror Symmetry, Mirror Map and Applications to
Complete Intersection Calabi-Yau Spaces'', Nucl. Phys. B433 (1995)
501 \semi P. Candelas, A. Font, S. Katz and
D.R. Morrison,``Mirror Symmetry for Two Parameter Models'',
Nucl. Phys. B429 (1994) 626.}.
Since we are now taking $F_g$ to correspond to certain
 exact computations in the full theory
we should be able to understand
their singularity from the viewpoint of nearly massless black holes.
This is what we will presently do.

In all the examples considered in \beret \holan \multi\ there was
a single conifold point $p$ (and codimension one in the 2
parameter family studied in \multi ).
  Let $z$ denote the period
of the vanishing cycle as we approach $p$.  In particular
$z(p)=0$.  The singularity that was encountered is of the form
$$F_1={-1\over 12}{\rm log}z\bar z +...$$
\eqn\sings{F_g={{\rm const.}\over z^{2g-2}}+...\qquad g>1}
The precise coefficient of singularity for $F_1$, ${-1\over 12}$,
was computed by fixing behaviour of $F_1$ at other boundaries
of moduli space where there was a better understanding of its behaviour.
It was thus quite mysterious that in all the four examples considered in
\holan\  and all the other examples including the 2 parameter
family ones studied in \multi ,
the coefficient of the conifold singularity came out to by
${-1/12}$ (there is an extra factor of 1/2 relative to that
in \holan ; as explained in \beret\ this comes from the $Z_2$ symmetry
of the torus).
Let us first consider the genus 1 term.  In this case \efac\ shows
that we are computing a term of the form $R^2$. In order to fix
the absolute normalizations of string one loop computation
 it turns out to be convenient to consider
a part of the interaction given by the index contraction leading
to the Euler characteristic density.  The rest
will be determined by $N=2$ supersymmetry.
The idea is to consider further compactification on a four manifold--
in this case in the computation of the string one-loop
partition function the term corresponding to the Euler-characteristic
would come from the sector where both left- and right-movers
have the same spin structure.  If the spin structure is odd, it clearly
gives the Euler characteristic $\chi$.  If we consider the sum of the 9
spin structures where they are  even on both left- and right-movers,
by supersymmetry one can show that we also get $\chi$.  Now, there is an
overall factor of $1/4$ coming from GSO projection factor.  So putting
all this together we find that
the precise string one loop normalization
is such that it gives
$$\delta_1 S={1/2\over 128\pi^2}\int F_1
\epsilon^{abij}\epsilon^{cdkl}R_{abcd}R_{ijkl} = {F_1\cdot \chi\over 2}$$
We are thus predicting a correction to the effective action
which near the conifold point is of the form
\eqn\spre{\delta_1 S={-\chi \over 24}{\rm log}z\bar z={-\chi \over 24}
 {\rm log}m^2}
Where would this come from in the effective theory where we have
a nearly massless black hole with mass $m$?  Counting the powers
of derivatives we see that it should come from a one loop
computation in the effective theory (which is also responsible
for giving the tree level string singularity as suggested
in \strom ).  Indeed it is easy
to see that such a term is generated at one loop.
  At one loop we would have to
compute
$${-1\over 2}{\rm STr}\ {\rm log}(-D^2+m^2)={-1\over 2}\int_\epsilon^{\infty}
{ds\over s} {\rm Tr}(-1)^F{\rm exp}[-s(-D^2+m^2)]$$
where $D^2$ is the Laplacian acting on the fields in the multiplet
and $(-1)^F$ is $\pm 1$ depending on whether we are dealing
with a boson or a fermion in the multiplet.
The term in this expansion proportional to  ${\rm log}\ m^2$ comes
from the order $s^0$ expansion of the heat kernel
${\rm Tr}(-1)^F{\rm exp}[-s(-D^2)]$.  The order $s^0$ term
can be expanded in the metric, and by dimension
counting it includes terms with four derivatives.
   Suppose
we have $N$ hypermultiplets.  Recall that a hypermultiplet
has 2 complex scalars and 2 Weyl fermions.  Thus we find a one
loop term of the form
$$\delta S={-N\ {\rm log}m^2\over 2}[4b_4(0,0)-b_4({1\over 2},0)
-b_4(0,{1\over 2})]$$
where $b_4(n,m)$ denotes the coefficient of heat kernel
expansion for a particle of spin $(n,m)$ under $SO(4)$.
These have been computed in the literature
\ref\seel{B.S. DeWitt,``Dynamical Theory of Groups and Fields'',
(Gordon and Breach, New York 1965)\semi
P.B. Gilkey, J. Diff. Geom. 10 (1975) 601\semi
S.M. Christensen and M.J.Duff, Nucl. Phys. B154 (1979) 340.}.
Again by $N=2$ supersymmetry
it suffices to consider the term proportional to Euler characteristic.
It turns out that
$$4b_4(0,0)-b_4({1\over 2},0)
-b_4(0,{1\over 2})={\chi \over 12}+...$$
We thus find for $N$ hypermultiplets of mass $m$ a term
in the effective action of the form
\eqn\ft{\delta S= {-N \chi \ {\rm log}m^2\over 24}}
Comparing this with the exact string theory computation
\spre\ we get agreement if $N=1$
as was required for the consistency of the
picture advocated in \strom .  This also `explains'
the mysterious factor of ${-1/12}$ appearing in all the
examples studied in \holan \multi.

Note that the special geometry of moduli of Calabi-Yau
manifold has a bearing on the coupling of $F^2$
as a function of moduli, whereas the one loop string computation
discussed above
has a bearing on the coupling of $R^2$.  In particular
the tree level amplitude is sensitive only to {\it charged}
states, whereas the one loop term, being a gravity effect,
takes into account {\it all} states.  In particular
in finding an agreement between the one loop and tree level counting
of massless states
we are finding evidence that there are no additional neutral fields
which are becoming massless at the conifold point\foot{ It could
happen, however, that we have extra massless hypermultiplets
and vector multiplets, equal in number, since the contribution
of vector multiplets turns out to be opposite in sign to
that of hypermultiplet in the coefficient of $R^2$.}.
It is amusing that also
for string-string duality in six dimensions in order to reproduce
the sigma-model
anomaly of heterotic string in the type IIA string, it
was found \vw\ that the tree level gives the $F^2$ part and
the one loop gives the $R^2$ part.

In the effective theory near the conifold
not only the nearly massless hypermultiplet generates
a one loop term proportional to $\chi$, but all hypermultiplets
do.
It is thus tempting to go further and identify $F_1$ with a partition
function for {\it all} BPS states:
\eqn\parti{F_1={-1\over 12}\sum_{BPS\ states}{\rm log}\ m_i^2}
where the above sum on the right hand side may need to be regularized
and where $m_i^2$ is given by the formula \masf .
This may not be precisely right: There
is a jumping phenomenon and so the BPS states do not necessarily
form a modular invariant subspace, however $F_1$ is modular invariant.
Moreover in the effective theory describing the vicinity of the conifold
 there may be other terms in the action proportional to $\chi$.
Another check on \parti\ is to study the anomaly equation it
satisfies \holan .
By considering moduli derivatives
 $\partial \bar \partial F_1$, if \parti\
were to
hold, we see using \masf\
that it would be proportional to $\partial \bar \partial
K$ where ${\rm exp}(-K)=\int_M \omega \wedge \bar \omega$ which according
to \holan\
is only
a piece of the anomaly satisfied by $F_1$.  In fact $F_1$ is given by
$$F_1={\rm log}\big[{\rm exp}[({3\over 2} +{h^{2,1}\over 2} +{\chi \over 24})K]
\ {\rm det} \ G^{-1/2}\ \big| f\big|^2\big]$$
where $f$ is a holomorphic function of moduli and $G=\partial \bar \partial K$
is the metric on the moduli space.  The extra non-holomorphicity
comes from non-holomorphicity of $G$.
But we also have to recall that in considering BPS states in \parti\
we have to take into account the massless fields corresponding
to the moduli.  The factor of ${\rm log}\ {\rm det}\ G^{-1/2}$ is exactly
right to come
from the normalization of the path integral for the massless moduli field.
At any rate, even if $F_1$ is not
precisely the partition function
of BPS states, I believe \parti\ should be very closely related to it,
and should be an important ingredient in fixing all the BPS states.
This is analogous to ideas suggested in \ref\fer{
S. Ferrara, C. Kounnas, D. Lust and F. Zwirner,``Duality-Invariant
Paritition Functions and Automorphic Superpotentials for (2,2)
String Compactifications'', Nucl. Phys. B365 (1991) 431.}\
in constructing automorphic forms on moduli space of Calabi-Yau
and relating it to sums over specific three-cycles.
To motivate it even further let us recall that $F_1$
computes a specific combination of Ray-Singer torsions on the Calabi-Yau
threefold \beret , i.e. a combination of logarithm of the determinant
of Laplacian acting on forms of the Calabi-Yau manifold.  Let us consider
an analogy:  Suppose we consider a one dimensional Calabi-Yau, i.e.
a torus. In that case the analog of the three-cycles on the threefold
are the one-cycles on the torus. Then the analog of candidates
for BPS states would
be one-cycles and the analog of the formula \parti\ is\foot{Summing
over primitive one-cycles will give the same result.}
$$F_1=-{\rm log}(det D^2)=\sum_{n,m}{\rm log} {\big| n+m\tau \big|^2
\over \tau_2} $$
which is a correct relation.
Thus in the threefold we would be suggesting a relation
between the particular combination of Ray-Singer torsions
appearing in \beret\ and a subspace of three-cycles in the
Calabi-Yau manifold corresponding to BPS solitons.
It should
be interesting to develop this further.  In particular it should
be interesting to see what happens if we restrict the sum in \parti\
to the space of vanishing cycles ${\cal V}$ discussed before.
Another interesting point to study would be the behavior
of $F_1$ for boundaries of moduli space which are infinitely
far away (analogous to infinite volume)
and study what it teaches us about the full spectrum of
the theory in that limit.

 It is easy to see by power counting that in order to obtain the higher genus
 superpotential terms computed in string theory,
we need higher loop corrections in the effective theory as well.
  In such a scenario near the conifold point
the powers of $m^2$ in the denominator in \sings\ would come from
infrared divergencies associated with a nearly massless
black hole.
Power counting
suggests that such a singularity could appear in a $g$-th
loop correction in the effective theory; for example having
a loop of nearly massless black holes with internal graviton
exchanges will give the correct power counting to reproduce
the effective action computed in string theory.

So far we have talked about compactifications of type II
strings on Calabi-Yau manifolds down to 4 dimensions.
If we instead consider type II strings compactified on $K3$ to
6 dimensions, there are superpotential terms that are
again captured by a topological string theory: The $N=4$
topological string\ \ref\berkv{N. Berkovits and C. Vafa,``N=4
Topological Strings'', HUTP-94/A018, KCL-TH-94-12.}\
 which is equivalent
to $N=2$ strings.  Just as the above superpotential
terms generated upon compactification on Calabi-Yau
is relevant in studying certain non-perturbative aspects
of string theory,
it is natural to expect that the $N=2$
string is important in a better understanding of heterotic-type IIA
duality in six dimensions. This question is currently under
investigation \ref\bov{N. Berkovits, H. Ooguri and C. Vafa,
Work in progress.}.

I would like to thank N. Berkovits, M. Bershadsky, S. Kachru,
 H. Ooguri, M.H. Sarmadi,
A. Strominger, E. Witten and S.-T. Yau for valuable discussions.

This research was supported in part by the NSF grant PHY-92-18167.

\listrefs

\end